\documentclass[conference]{IEEEtran}
\usepackage{algorithm}
\usepackage{algorithmicx}
\usepackage{algpseudocode}
\IEEEoverridecommandlockouts
\usepackage{cite}
\usepackage{amsmath,amssymb,amsfonts}
\usepackage{booktabs}
\usepackage{multirow} 
\usepackage{graphicx}
\usepackage{textcomp}
\usepackage{xcolor}
\usepackage{upgreek}
\usepackage{cleveref}
\crefrangeformat{equation}{#3(#1)#4-#5(#2)#6}
\usepackage[top=0.73in, bottom=1in, left=0.68in, right=0.613in]{geometry}

\def\BibTeX{{\rm B\kern-.05em{\sc i\kern-.025em b}\kern-.08emT\kern-.1667em\lower.7ex\hbox{E}\kern-.125emX}}
\usepackage{url}
\begin{document}

\title{Aerial IRS with Robotic Anchoring Capabilities: A Novel Way for Adaptive Coverage Enhancement }

\author{
    \IEEEauthorblockN{Xinyuan Wu
    and  Vasilis Friderikos}
    \IEEEauthorblockA{{Centre for Telecommunications Research, Department of Engineering,}
    \IEEEauthorblockA{King's College London, London WC2R 2LS, UK.}}
    \IEEEauthorblockA{E-mail:\{xinyuan.1.wu, vasilis.friderikos\}@kcl.ac.uk}
}

\maketitle

\begin{abstract}
It is widely accepted that integrating intelligent reflecting surfaces (IRSs) with unmanned aerial vehicles (UAV) or drones can assist wireless networks in improving network coverage and end user  Quality of Service (QoS). However, the critical constrain of drones is their very limited  hovering/flying time. In this paper we propose the concept of robotic aerial IRSs (RA-IRSs), which are in essence drones that in addition to IRS embed an anchoring mechanism that allows them to grasp in an energy neutral manner at tall urban landforms such as lampposts. By doing so, RA-IRSs can completely eliminate the flying/hovering energy consumption and can offer service for multiple hours or even days (something not possible with UAV-mounted IRSs). Using that property we show how RA-IRS can increase network performance by changing their anchoring location to follow the spatio-temporal traffic demand. The proposed methodology, developed through Integer Linear Programming (ILP) formulations offers a significant  Signal-to-Noise (SNR) gain in highly heterogeneous regions in terms of traffic demand compared to fixed IRS; hence, addressing urban coverage discrepancies effectively. Numerical simulations validate the superiority of RA-IRSs over fixed terrestrial IRSs in terms of traffic serviceability, sustaining more than 2 times the traffic demand in areas experiencing high heterogeneity, emphasizing their adaptability in improving coverage and QoS in complex urban terrains. 

\end{abstract}


\section{Introduction}

\IEEEPARstart{T}{he
} amalgamation of Intelligent Reflecting Surfaces (IRS) and Unmanned Aerial Vehicles (UAVs) is recognized as a way to further increase overall radio access network performance in diverse settings \cite{intro3}. More specifically, the inherent capabilities of IRS in adaptively controlling electromagnetic waves, assist in alleviating detrimental impacts due to wireless channel characteristics. Simultaneously, UAVs exploit their superior mobility and elevated positioning to establish optimized Line-of-Sight (LoS) conditions \cite{intro4}. This integration of those two components has recently attracted significant attention and is considered as another important technique to augment wireless networks. This include, inter alia, enhanced network connectivity \cite{intro4}, architecting energy-efficient communication frameworks \cite{intro5}, enabling intelligent and secure vehicle to infrastructure communication \cite{intro6}, and realizing integrated sensing and communication, such as advanced sensing and data acquisition \cite{intro7}. The aforementioned diverse set of applications underscore the significant potential of the IRS-UAV integration in propelling wireless communication technologies forward \cite{intro8}.

Nevertheless, the inherent energy limitations of UAVs pose substantial challenges to the seamless integration of UAV and IRS. To address these issues, several innovative solutions have been proposed. One such solution is the integration of solar panels on UAVs, serving as a renewable energy source to extend operational duration and mitigate energy constraints \cite{intro9}. Another approach focuses on the utilization of terrestrial IRS to optimize energy consumption and enhance communication performance in UAV-centric networks \cite{intro4}. Additionally, the development of tethered structures has been explored to facilitate the provision of aerial MEC services in areas devoid of infrastructure, offering a viable solution to infrastructure-related limitations \cite{intro10}. However, such solutions to energy constraints necessitate trade-offs in terms of flexibility or hardware costs, requiring careful consideration of the balance between benefits and drawbacks.
\begin{figure}[tb]
    \centering
    \includegraphics[width=0.75\linewidth]{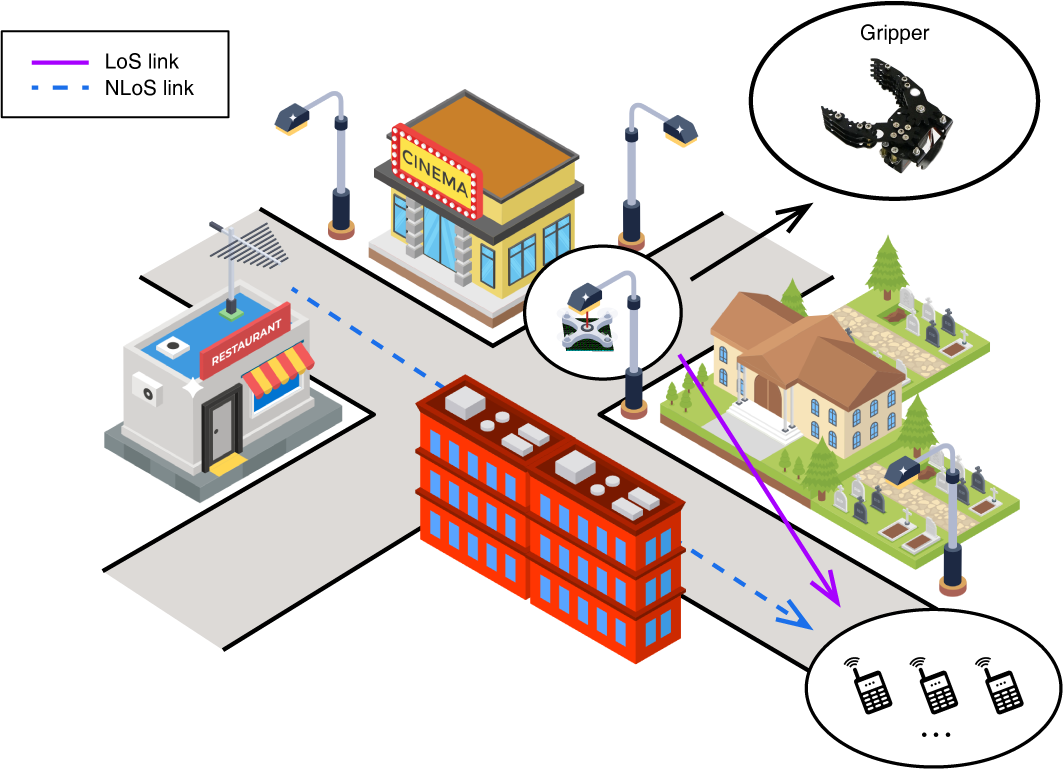}
    \caption{An overview of RA-IRS-assisted B5G mmWave microcell scenario.}
    \label{fig:1}
\end{figure}

Inspired by the Robotic Aerial Base Stations (RABS) concept \cite{gripper}, this paper proposes a novel concept termed Robotic Aerial Intelligent Surface (RA-IRS), achieved by integrating a robotic gripper with Aerial Intelligent Reflecting Surfaces (AIRS). More specifically, RA-IRS is able to grasp using energy neutral robotic anchoring mechanisms at tall urban landforms (such as street lights) and is  equipped with a passive IRS to extend cell coverage in areas experiencing low QoS such as coverage holes. RA-IRS shall be deployed in street canyons within microcells, where intricate environments often result in signal blockages. The incorporation of a gripper allows RA-IRS to anchor into different urban street furniture, such as lampposts for example, and significantly mitigate the power consumption for hovering/flying whilst  establishing virtual Line-of-Sight LoS paths to low QoS areas, thereby catering to regions with higher traffic demands over extended duration. Given the limited physical size of the IRS deployed on the UAV, we assume that each RA-IRS can serve only one area. Given the abundant spectral resources of the mmWave band and the use of orthogonal resource blocks, interference considerations are significantly mitigated in our context. 
The Manhattan grid model is employed to partition the coverage area of an urban micro base station into uniform grids, each represented by a central user depicting the mean QoS, and intra-grid quality variations are minimized due to the compact nature of each grid\footnote{ In the context of the Manhattan model, the term `grid' is synonymous with `area' in a microcell and the two terms are used interchangeably hereafter}. 


\section{System Model}
\subsubsection{IRS Design}
\indent The IRS to be integrated on UAVs is based on an Integrated-architecture, offering marked advantages in energy efficiency, electromagnetic interference resistance, and compactness compared to the conventional FPGA-based architecture \cite{HF0}. Typically, IRS elements adopt a $\lambda/2$ spacing to reduce spatial correlation \cite{power0}. Letting $N_{\mathrm r}$ represent the number of elements along each dimension of a square-shaped IRS, the Fraunhofer distance, $D_{F}$, can be expressed as $D_{F} = \frac{\lambda}{2}N_{\mathrm r}^{2}
$\cite{farfield}. This distance $D_{F}$ should not exceed the minimum IRS-to-transceiver distance, $D_{\mathrm{min}}$. Furthermore, with a 2-bit discrete coding scheme which provides near-optimal performance and enhanced energy efficiency \cite{numofbits1}, it will be beneficial if $N_{\mathrm r}$ is a multiple of 4 to mitigate potential discontinuity in phase gradient \cite{multiple4}.
\subsubsection{RA-IRS Energy Model}
The energy consumption of RA-IRS comprises three components: the propulsion energy incurred when RA-IRS moves between different candidate locations; the grasping energy of the gripper when RA-IRS is suspended from aerial platforms; and the reflecting energy from the IRS on RA-IRS, which facilitates anomalous reflection to provide a virtual-LoS path.
\paragraph{Propulsion Energy}
The propulsion power, $P_{\text{fly}}$ is dependent on UAV's flight velocity $v$ and weight $m_{\mathrm{tot}}$ \cite{flypower}. Given that the candidate locations are assumed to be at the same altitude, its movement is only considered in the horizontal plane. For two such locations with coordinates $L_{p}$ and $L_{q}$, the propulsion energy is given by 
$E_{\text{fly}} = P_{\text{fly}} \times \frac{||L_{p} - L_{q}||}{v_{\mathrm{fly}}}$ where $||\cdot||$ denotes the L2-norm.
\paragraph{Grasping Energy}
The proposed RA-IRS, designed for attachment to aerial platforms such as lampposts, prefers an electromagnetic solenoid-based gripper \cite{gripper}. The gripper shall operate continuously, with the exception of the brief RA-IRS propulsion intervals, thus the grasping energy can be upper-bounded by $E_\mathrm{{grasp}} = P_{\mathrm{grasp}} \times T$. Here, $P_{\mathrm{grasp}}$ denotes the power consumption of the gripper, influenced by its physical characteristics, while $T$ represents the total service duration of RA-IRS.
\paragraph{Reflecting Energy}
The anomalous reflection is facilitated through the previously mentioned Integrated-architecture 2-bit discrete coding IRS. Its power consumption consists of both static and dynamic components. Specifically, \(P_{\mathrm{IRS}} = P_{\mathrm{st}}^{\mathrm{tot}} + P_{\mathrm{dyn}}^{\mathrm{tot}}\), which includes energy drawn from control chips and impedance-adjusting diodes \cite{power0}. Analogous to the gripper, the IRS remains operational except during the propulsion intervals. Hence, $E_{\mathrm{ref}} = P_{\mathrm{IRS}} \times T$.

\subsubsection{Network and Transmission Model}
In a mmWave microcell, assuming the BS and user terminals are each fitted with a single antenna \cite{model2}. Denote the base station's position as $B = (b_{x}, b_{y})$. The set of grid center positions is given by $\mathcal{U} = \{ U_1, U_2, \dots, U_I \}$, where $i\in\{1,2,\cdots,I\}\triangleq\mathcal{I}$ as the index. The distance from any grid center $U_{i}$ to the base station, considering height difference $H_{1}$ between each UT and BS, is computed as $\mathcal{L}_i = \sqrt{||U_{i}-B||^{2} + H_{1}^{2}}$. For a set of proposed candidate locations $\mathcal{P}= \{ P_1, P_2, \dots, P_J \}$ that is indexed by $j\in\{1,2,\cdots,J\}\triangleq\mathcal{J}$. The distances to the base station and from the IRS to the user terminal are respectively, $\mathcal{R}_j = \sqrt{||P_{j}-B||^{2} + H_{2}^{2}}$ and $\mathcal{D}_{ij} = \sqrt{||U_{i}-P_{j}||^{2} + H_{3}^{2}}$. Here, $H_{2}$ and $H_{3}$ denote the height differences from the landform to the base station and user terminal, respectively.

In the absence of detailed geographical data or ray tracing references, we compute the LoS probability for each grid using a distance-dependent empirical formula \cite{3GPP}:
$\mathrm{Pr}_{\mathrm{LoS}}(d) = \frac{18}{d} + \exp\left(-\frac{d}{36}\right)(1-\frac{18}{d})\ \text{if } d \geq 18\ \textrm{m} $, and 1 otherwise.
Here d is the 2-D distance of each BS-UT link.
For each grid $U_i$, if $ r_i \in [0, 1] $ is a uniformly distributed random number, the grids lacking a LoS connection in a given trial are denoted by:
$\mathcal{S} = \{ i \mid \mathrm{Pr}_{\mathrm{LoS}}(d_{i}) > r_i,\ d_{i} = ||U_{i}-B||, \ U_{i} \in \mathcal{U},\ i\in\mathcal{I}\}.$

Small scale fading between each BS-UT direct link is supposed to follow Rician distribution, characterized by factor $\mathcal{K}$. For grids in $\mathcal{S}$, which indicates NLoS conditions, $\mathcal{K}_{i} = 0$ simulates Rayleigh fading. For LoS grids, $\mathcal{K}_{i} = K_{d} $, where $K_{d}$ is the power ratio of the LoS path to multipath components. The channel coefficient is:
\begin{equation}
h_{d}^{i}=\sqrt{\frac{\mathcal{K}_{i}}{1+\mathcal{K}_{i}}}\bar{h}_{d}^{i}+\sqrt{\frac{1}{1+\mathcal{K}_{i}}}\tilde{h}_{d}^{i},
\end{equation}
where \(\bar{h}_{d}^{i}\) and \(\tilde{h}_{d}^{i}\sim\mathcal{CN}(0,1)\) represent the LoS and multipath NLoS components, respectively\cite{model2}. 

Large scale path loss is meanwhile described as:
$PL_{d}^{i}=M_{i}\cdot A_{d}{\mathcal{L}_{i}}^{-\eta_{1}}+(1-M_{i})\cdot A_{d}{\mathcal{L}_{i}}^{-\eta_{2}}$, where $M_{i}=1$ for $i\in\mathcal{S}$ and 0 otherwise. Here, $A_{d}$ signifies the reference path loss at a unit distance, while $\eta_{1}$ and $\eta_{2}$ denote the PLE for terrestrial LoS and NLoS links, respectively \cite{tutorial}. 

Under the condition of slow and flat fading channels, the average received SNR associated with each grid is denoted by 
\begin{equation}
\gamma_{d}^{i}=PL_{d}^{i}\left[\mathrm{E}(|h_{d}^{i}|^{2})\right]\frac{P_{t}}{N_{0}}=PL_{d}^{i}\frac{P_{t}}{N_{0}},
\end{equation} 
where $\mathrm{E}\left(|h_{d}^{i}|^{2}\right)$ is normalized to 1, and $P_{t}$ and $N_{0}$ respectively represents total transmitted power and total noise power \cite{model2}.

\indent To better identify the grids that critically require RA-IRS support, we further narrow down the grids based on their received SNR. We define a set $\mathcal{Q} $ given by
$
\mathcal{Q} = \{ i \mid i \in \mathcal{S},\ \gamma_{d}^{i} < \gamma_{\text{th}} \},
$ where $\gamma_{\text{th}}$ is a predetermined SNR threshold. 

\indent Given the optical visibility of the virtual-LoS link facilitated by RA-IRS, we model the BS-IRS-UT link as undergoing cascaded Rician fading. Consider a sub-path $l$ associated with a specific unit element on the IRS, denoted as $l \in \left\{ 1, 2, \ldots, N_{\mathrm{IRS}} \right\}$, where $N_{\mathrm{IRS}} = N_{\mathrm r}^{2}$ represents the total number of elements on the IRS. The fading coefficients corresponding to the BS-IRS and IRS-UT links are defined as $h_{c,1}^{l} = \alpha_{i}e^{-j\theta_{i}}$ and $h_{c,2}^{l} = \beta_{i}e^{-j\phi_{i}}$, respectively \cite{model2}. Each unit element on the RA-IRS is characterized by a reflection coefficient given by $\Gamma e^{j\varphi_{l}}$, where $\Gamma$ ideally reaches 1, and $e^{{\varphi_{l}}}$ denotes the controllable phase shift. By possessing perfect channel state information at the IRS, the optimal phase shift for each element can be selected as $\varphi_{l} = \theta_{l} + \phi_{l}$, thereby nullifying potential phase discrepancies \cite{tutorial}.

\indent Assume that the path losses of the $N_{\mathrm{IRS}}$ reflected paths are identical \cite{tutorial}, the path-loss of the virtual-LoS link can be denoted by $PL_{c}^{ij} = A_{t}\mathcal{R}_{j}^{-\eta_{3}} \cdot A_{r}\mathcal{D}_{ij}^{-\eta_{3}}$, where $j$ indicates the chosen candidate location, and $A_{t}$ and $A_{r}$ respectively represent the reference path losses for BS-IRS and IRS-UT links. The large scale PLE $\eta_{3}$ is notably less than that in the terrestrial NLoS scenario \cite{model2}.

\indent Leveraging the Central-Limit Theorem for a sufficiently large $N_{\mathrm{IRS}}$, the average end-to-end SNR of the cascading link through RA-IRS from candidate location $j$ to grid $i$ is given by 
\begin{equation}
\gamma_{c}^{ij} = PL^{ij}_{c}\left\{N_{\mathrm{IRS}}+\frac{\pi^2}{16}\frac{\left(N_{\mathrm{IRS}}^{2}-N_{\mathrm{IRS}}\right)}{\left[{\sqrt{\frac{1}{1+\mathcal{K}_{c}}}}\frac{1}{L_{1/2}(-\mathcal{K}_{c})}\right]^{4}} \right\}\frac{P_{t}}{N_{0}},
\end{equation}
with $L_{1/2}(\cdot)$ being the Laguerre polynomial of degree $1/2$\cite{model2,model3}.
 
We define the SNR ratio, \( \mathcal{G}_{ij} \), as the fraction of the aggregated SNR to that of the direct link, given by \( \mathcal{G}_{ij}=\frac{\gamma^{i}_{d} + \gamma^{ij}_{c}}{\gamma^{i}_{d}} \). This ratio evaluates the impact of RA-IRS deployment at a specific candidate location \( j \) for grid \( i \). Given the limited on-board battery of UAVs and the sporadic nature of traffic demand, our analysis is confined to specific time slots within a subset denoted as $\mathcal{T}$. The traffic demand of a specific grid $i$ at time slot $t$ follows log-normal distribution is represented by $\mathcal{F}_{i}(t)$ \cite{traffic}. With a predetermined traffic threshold $\mathcal{F}_{th}$ below which traffic demand is supposed to be negligible, the traffic-dependent SNR gain can be defined as:
\begin{equation}
\underset{i \in \mathcal{Q},t \in\mathcal{T},j\in\mathcal{J}}{\mathcal{G}_{tij}} = \begin{cases}
 \mathcal{G}_{ij} & \text{if } \mathcal{F}_{i}(t) \geq \mathcal{F}_{th}  \\
1 & \text{otherwise}
\end{cases}.
\end{equation}
This formulation ensures that grids with lower traffic demand are not given priority, with the overarching objective being to maximize the traffic-dependent gain over both area and time.

\section{Problem Formulation}
\subsubsection{Joint Placement and Serving-Area Selection}
We aim to optimize the SNR gain in regions experiencing low QoS throughout the entire operational period of $M$ deployed RA-IRSs, where $M\leq|\mathcal{Q}|$. We are doing so by formulating a strategy that jointly selects serving areas and determines corresponding placement of each RA-IRS at every time epoch. To this end, we introduce a binary variable $x_{tij} \in \{0, 1\}$ where $x_{tij} = 1$ denotes that one RA-IRS is allocated to serve grid $i$ and anchor at candidate location $j$ during time epoch $t$, and $x_{tij} = 0$ otherwise. The objective is to maximize the overall SNR gain across the target areas, and the proposed optimization problem can be formulated as follows,
\begin{align}
\text{(P1):} \ \max_{x_{tij}} &\  \frac{1}{|\mathcal{T}||\mathcal{Q}|}\sum_{t\in\mathcal{T}} \sum_{i\in\mathcal{Q}} \sum_{j\in\mathcal{J}} \left[x_{tij} \cdot \mathcal{G}_{tij}+\underbrace{(1-x_{tij})\cdot{1}}_{\text{Gain of unserved areas}}\right] \label{5a}\tag{5a} \\
\text{s.t.} \ & \sum_{i\in\mathcal{Q}} \sum_{j\in\mathcal{J}} x_{tij} = M, \quad \forall t\in\mathcal{T} \label{5b}\tag{5b} \\
& \sum_{i\in\mathcal{Q}} x_{tij} \leq 1, \quad \forall t\in\mathcal{T}, \forall j\in\mathcal{J} \label{5c}\tag{5c} \\
& \sum_{j\in\mathcal{J}} x_{tij} \leq 1, \quad \forall t\in\mathcal{T}, \forall i\in\mathcal{Q} \label{5d}\tag{5d} \\
& x_{tij} \in \{0,1\}, \quad \forall t\in\mathcal{T}, \forall i\in\mathcal{Q}, \forall j\in\mathcal{J} \label{5e}\tag{5e},
\end{align}
note that grids not selected for service are presumed to obtain a unit SNR gain. Here, constraint (5b) ensures the placement of exactly $M$ RA-IRSs at each time epoch, constraints (5c) and (5d) guarantee that each RA-IRS serves only one area and each area is serviced by a single RA-IRS, respectively. 

\begin{itemize}
    \item[(a)] \textbf{Terrestrial IRS:} 
    As a point of comparison, we define the scenario for the Terrestrial IRS as given in (P1.a). The objective function (\ref{5a}) and constraints \crefrange{5b}{5e} are identical to those in (P1). The newly introduced constraint (\ref{5f}) indicates that, due to the lack of mobility, the Terrestrial IRS cannot switch its anchoring candidate location or the area it services across different time epochs. Consequently, the optimization to maximize the SNR Gain is only implemented in the first time epoch.
    \begin{align}
    \text{(P1.a):} \ &\ (\ref{5a}) \notag\\
    \text{s.t.} \ &\crefrange{5b}{5e}\notag\\
    & x_{tij} = x_{1ij}, \quad \forall t\geq2, \forall i\in\mathcal{Q},\forall j\in\mathcal{J} \label{5f}\tag{5f}
    \end{align}
    
    \item[(b)] \textbf{Random Sampling:}
    Furthermore, we define a baseline (P1.b). Its scenario and constraints are similar to (P1.a), but the placement and area selection of the Terrestrial IRSs in the first time epoch is conducted through random sampling.
\end{itemize}

\subsubsection{Trajectory Planning}
Once the solution of P1 is obtained, we proceed to consider the trajectory planning of RA-IRSs to determine $M$ paths over the serving time, with the aim of minimizing the cumulative traveling distances and, consequently, reducing the overall energy consumption. Based on the solution of P1 the area to be served by a RA-IRS is fixed at a given time epoch. Thus, we represent the solution of P1 as $\mathbf{X} \in \mathbb{R}^{T \times M}$, where each element is denoted by $x_{tm},\ 1 \leq t \leq T,\ 1 \leq m \leq M$. Here, $m$ represents the index of the location in the solution matrix $\mathbf{X}$ at time $t$, indicating the placement of RA-IRSs at different locations at each time epoch.

\indent Next, we introduce a distance matrix $\mathbf{V} \in \mathbb{R}^{(T-1) \times M \times M}$, with element $v_{tjk}$ representing, without loss of generality, the Euclidean distance between locations indexed by $j$ and $k$ at consecutive time slots $t$ and $t+1$. Here, $j$ and $k$ are indices representing the locations in the solution matrix $\mathbf{X}$ at times $t$ and $t+1$ respectively, indicating the transition of RA-IRSs between different locations across consecutive time epochs. $j_{0}$ and $k_{0}$ serve as two pseudo-indices both pointing to the location of the base station.

\indent Based on the aforementioned, we define a binary variable $y_{tjk}$ to indicate whether a RA-IRS travels from the location indexed by $j$ at time $t$ to the location indexed by $k$ at time $t+1$. To prevent collisions, constraints (6b) and (6c) are imposed, ensuring that each location is occupied by exactly one RA-IRS per time slot and that each RA-IRS moves from its current location to exactly one new location per time slot.

The ILP assignment problem P2 is formulated as follows:
\begin{align}
\text{(P2):}\ \min_{y_{tjk}} &\  \underbrace{\sum_{k=1}^{M}v_{0j_{0}k}}_{\text{before first slot}}+\sum_{t=1}^{T-1} \sum_{j=1}^{M}\sum_{k=1}^{M} y_{tjk} \cdot v_{tjk}+\underbrace{\sum_{j=1}^{M}v_{Tjk_{0}}}_{\text{after final slot}}\tag{6a}\\
\text{s.t.}\ & \sum_{j=1}^{M} y_{tjk} = 1, \quad \ 1 \leq t \leq T-1,\ 1 \leq k \leq M \tag{6b}\\
&  \sum_{k=1}^{M} y_{tjk} = 1, \quad\ 1 \leq t \leq T-1,\  1 \leq j \leq M \tag{6c}\\
& y_{tjk} \in \{0,1\}, \quad 1 \leq t \leq T-1,\ 1 \leq j,k \leq M \tag{6d}.
\end{align}

\begin{algorithm}
\centering  
\begin{minipage}{.45\textwidth}  
\caption{(P1.b): }
\begin{algorithmic}[1]
\State \textbf{Given parameters}
\State Initialize $x_{tij}$ as zeros with dimensions $(|\mathcal{T}|,|\mathcal{Q}|,|\mathcal{J}|)$
\State Set maxIterations
\For {iter = 1 to maxIterations}
    \State $x_{\text{temp}} \gets$ zeros with dimensions $(|\mathcal{Q}|, |\mathcal{J}|)$
    \State Randomly set $M$ entries of $x_{\text{temp}}$ to 1
    \State Check constraints \crefrange{5c}{5d}
    \If{(\ref{5c}) and (\ref{5d}) are met}
        \State $x_{1ij} \gets x_{\text{temp}}$
        \State \textbf{break}
    \EndIf
    \If{iter is maxIterations}
        \State \textbf{Terminate}
    \EndIf
\EndFor
\For{$t = 2$ to $T$}
    \State $x_{tij} \gets x_{1ij}$
\EndFor
\State Compute average gain:
\State $\frac{1}{|\mathcal{T}||\mathcal{Q}|}\sum_{t\in\mathcal{T}} \sum_{i\in\mathcal{Q}} \sum_{j\in\mathcal{J}} \left[x_{tij} \cdot \mathcal{G}_{tij}+(1-x_{tij})\cdot{1}\right]$
\end{algorithmic}
\end{minipage}
\end{algorithm}

Notably, the following two processes are incorporated in the strategy. Initially, before the first time slot, all RA-IRSs depart from the base station. Subsequently, after the final time slot, they return to the base station.

\section{Numerical Investigations}
We consider an urban microcell environment covering an area of $160 \times 160\ \mathrm{m^2}$, divided into a $9 \times 9$ grid, each being a 20 m side length square to represent typical street widths. A base station is located at the center of this grid area, using a carrier frequency of 28 GHz \cite{model2}. Uniformly spaced M3 category street lights are positioned along the streets at a height of $h_{\text{pole}}=12$ m on both sides, aligned with the street width \cite{pole1,pole2}. The number of available RA-IRSs $M$ is set to be 10 and all RA-IRSs are assumed to have identical physical parameters, as detailed in Table \ref{tab:RA-IRS}.
\begin{table}[tb]
    \centering
    \caption{Physical Parameters of RA-IRS}
    \begin{tabular}{l|c|l|c}
        \toprule
        \textbf{Parameter} & \textbf{Value} & \textbf{Parameter} & \textbf{Value} \\
        \hline\hline
       $m_{\mathrm{IRS}}$ & $0.1\ \mathrm{kg}$\cite{mass}&$m_{\mathrm{UAV}}$  & $4\ \mathrm{kg}$\cite{flypower1} \\
       $m_\mathrm{gripper}$ & $0.4\ \mathrm{kg}$\cite{gripper} & 
        $P_{\mathrm{IRS}}$ & $0.9\ \mathrm{W}$\cite{HF,power0} \\$P_{\mathrm{grasp}}$&$10\ \mathrm{W}$\cite{gripper1}&
         $P_{\mathrm{fly}}$ & $253.6\ \mathrm{W}$\cite{flypower}\\ $\ \ v_{\mathrm{fly}}$ & $10\ \mathrm{m/s}$\cite{flypower1}&$E_\mathrm{Battery}$&$799200\ \mathrm{J}$\cite{flypower1}\\
        \bottomrule
    \end{tabular}
    \label{tab:RA-IRS}
\end{table}

The minimum distance between the IRS and transceiver is attained when a mobile user is directly below the lamppost where the RA-IRS is anchored, resulting in $D_{\mathrm{min}}=10.5$ m. This choice of parameters leads to an IRS comprising $N_{\mathrm{IRS}}=48\times 48=2304$ elements. The corresponding far-field threshold, $D_{F}$, is set at 10.75 m, which is acceptable since users generally maintain a certain distance from lampposts. IRS with such dimension shall suffice to relax the aforementioned far-field constraint beyond which plane-wave assumption can be applied.

The urban traffic demand can be characterized by a log-normal distribution with a time-varied mean and a scenario-dependent standard deviation. Based on the reference \cite{traffic1}, the mean value is selected as 702 $\mathrm{Mbps/km^{2}}$, with fluctuations observed to be 20\% below to 40\% above the average across different time epochs. The standard deviation is contingent upon the type of region and is selected as 2.8 in the simulation unless specified otherwise \cite{traffic}. Consequently, the traffic threshold $\mathcal{F}_{th}(t)$ is defined as 1\% of the mean traffic demand generated at time $t$. A comprehensive summary of all simulation parameters is provided in Table \ref{tab:parameters}.
\begin{table}[tb]
    \centering
    \caption{Simulation Parameters}
    \begin{tabular}{l|c|l|c}
        \toprule
        \textbf{Parameter} & \textbf{Value} & \textbf{Parameter} & \textbf{Value} \\
        \hline\hline
       $H_{1},\ H_{2},\ H_{3}$&$8.5,\ 2,\ 10.5\ \mathrm{m}$\cite{3GPP}&$f_{c}$&$28\ \mathrm{GHz}$\\
       $N_{\mathrm{IRS}}$ & $48\times48$ & $\mathcal{K}_{c},\ \mathcal{K}_{d}$ & $10\ \mathrm{dB}$\cite{model2} \\
        $\eta_{1},\eta_{2}$ & $2.1,\ 3.17$\cite{model2} &$\eta_{3}$&$2.4$\cite{ple3} \\
        $A_{t},\ A_{r}$ & $-56.38\ \mathrm{dB}$\cite{model2}&$A_{d}$ & $-61.38\ \mathrm{dB}$\cite{model2}\\
        $P_{t}$ & $37\ \mathrm{dBm}$\cite{model2}&$N_{0}$ & $-95\ \mathrm{dBm}$\cite{model2}\\
        $\gamma_{th}$&$10\ \mathrm{dB}$&$\sigma$&$2.8$\cite{traffic1}\\
        \bottomrule
    \end{tabular}
    \label{tab:parameters}
\end{table}

\begin{figure}[tb]
    \centering
    \includegraphics[width=0.75\linewidth]{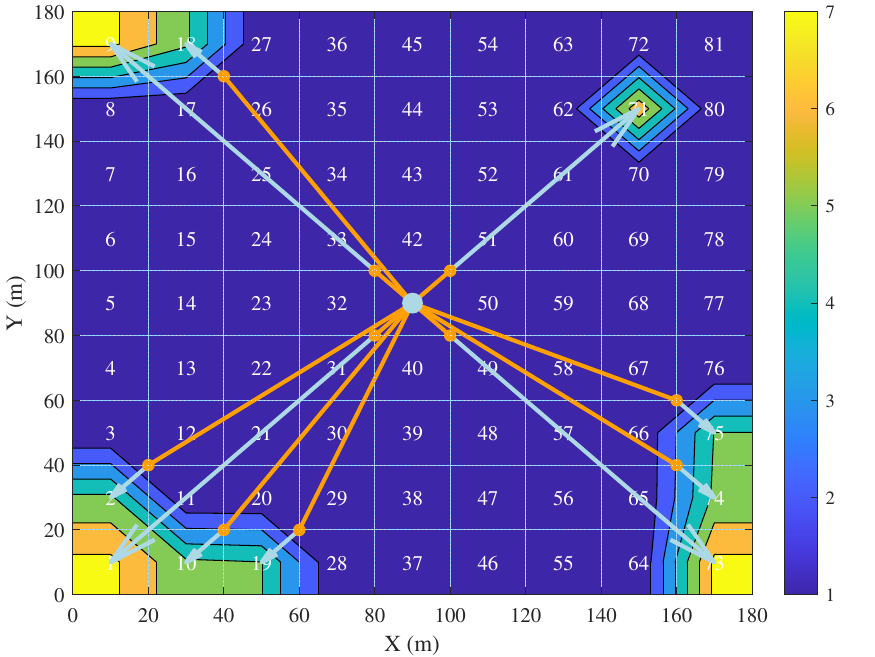}
    \caption{Placement and Area Selection of M=10 RA-IRSs at 1pm of a trial.}
    \label{fig:2}
\end{figure}

The solution to problem P1 offers a practical methodology for determining the optimal locations for deploying RA-IRSs and the appropriate times to assist specific areas based on the evolving spatio-temporal traffic variations. Fig. \ref{fig:2} depicts an illustrative example of traffic demand at 1 pm. In this figure, orange lines represent the connections between the base station (BS) and the robotic IRSs, while blue lines illustrate the links between the robotic IRSs and user terminals (UTs), with the four vertices of each grid serving as potential candidate locations for RA-IRSs. These connections together establish cascading virtual-LoS links. A key observation from the figure is the strategic positioning of robotic IRSs could be on either side of the transceiver, aligning with the findings presented in \cite{tutorial}. Furthermore, priority is given to servicing the most distant grids experiencing NLoS conditions due to the substantial improvements achievable by leveraging RA-IRSs, particularly in areas characterized by lower SNR.

\begin{figure}[tb]
    \centering
    \includegraphics[width=0.75\linewidth]{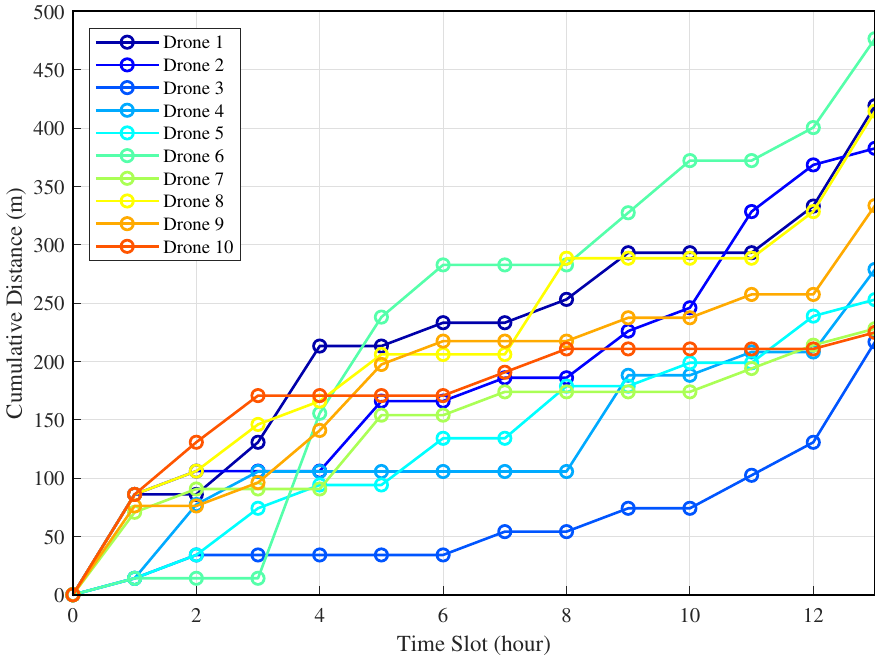}
    \caption{Cumulative Distance for M=10 RA-IRSs over Time of a trial.}
    \label{fig:3}
\end{figure}

In solving problem P2, optimal trajectories can be identified that minimize the flying distance, thereby reducing energy consumption. A visual representation of the cumulative flying distances for each RA-IRS over time during a trial is provided in Fig. \ref{fig:3}. Note that these trajectories may vary due to traffic conditions. However, given a total serving time of 12 hours with hourly spacing and adhering to the parameters listed in Table \ref{tab:RA-IRS}, the on-board battery, excluding that used for grasping and reflecting, allows each RA-IRS to fly for more than 12 km, which effectively addresses and mitigates concerns related to energy constraints.

\begin{figure}[htbp]
    \centering
    \begin{minipage}{.35\textwidth}
        \centering
        \includegraphics[ width=1\linewidth]{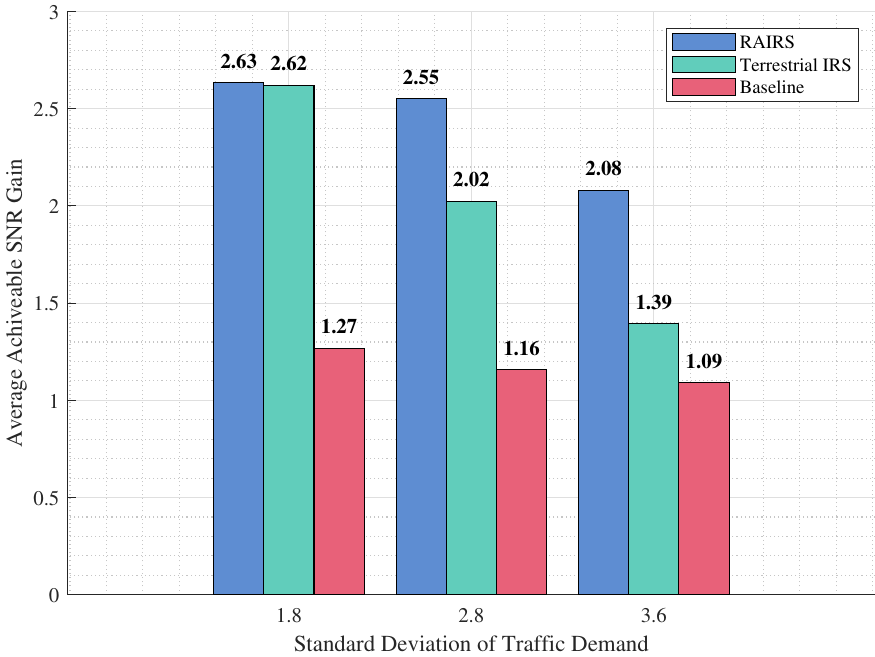}
        \caption{Comparison of Traffic-Aware SNR Gain vs. Traffic Heterogeneity.}
        \label{fig:4}
    \end{minipage}
    \hfill 
    \begin{minipage}{.35\textwidth}
        \centering
        \includegraphics[width=1\linewidth]{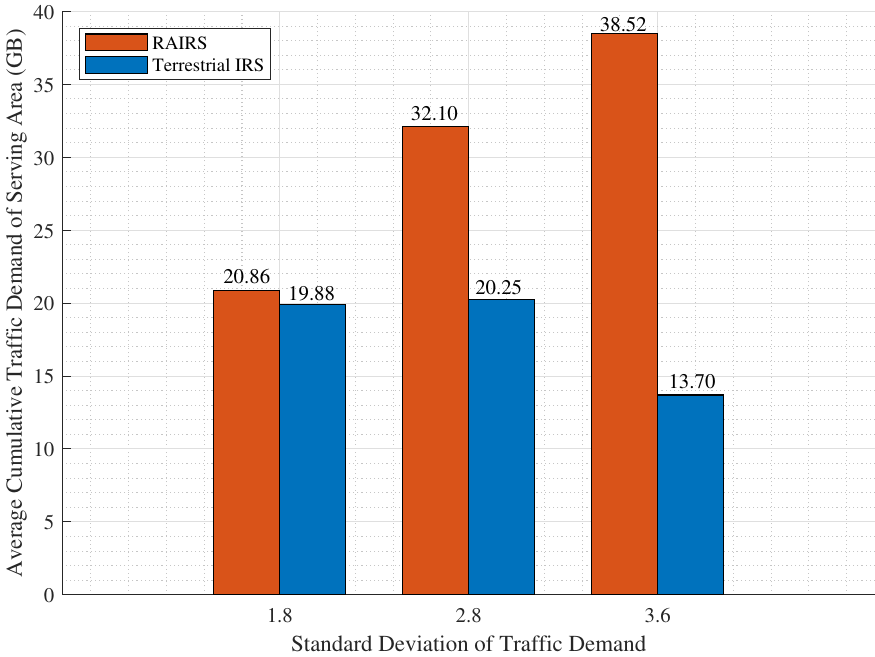}
        \caption{Comparison of Serving Traffic Demand vs. Traffic Heterogeneity.}
        \label{fig:5}
    \end{minipage}
\end{figure}

Figure \ref{fig:4} presents a comparative study between the advanced RA-IRS and the traditional terrestrial IRS under varying traffic standard deviations, specifically for \(\sigma = 1.8, 2.8,\) and \(3.6\). For bench-marking, a random sampling scenario serves as the baseline. The findings affirm that strategically optimizing the deployment of a sensible amount of IRSs can amplify the average gains by over two-fold, attributed mainly to superior coverage. Moreover, in situations characterized by high traffic heterogeneity, the RA-IRS model distinctly surpasses its terrestrial counterpart by $50\%$, thanks to its inherent mobility and comprehensive  coverage. It is crucial to underline that an increment in $\sigma$ is inversely related to the achievable traffic-dependent SNR Gain. This pattern emerges as there is an infrequent selection of zones capable of attaining higher SNR Gain, to accommodate areas with more intense traffic demands. This effect of pronounced traffic variations is more contained at lower $\sigma$ values.

Furthermore, Figure \ref{fig:5} depict the effect of the traffic variation in the serving zones for both terrestrial and RA-IRS configurations. At a $\sigma$ of 1.8, the RA-IRS system exhibits near-consistent traffic demands, overshadowing its terrestrial counterpart, signifying negligible traffic oscillations. However, with increased values of $\sigma$, these traffic disparities become more evident. In such cases, RA-IRS adeptly addresses zones witnessing a spike in traffic influx. Considering the innate traffic heterogeneity typical of urban street canyons, the RA-IRS approach is poised to present a notable edge over fixed terrestrial IRS, particularly concerning traffic-aware coverage enhancement.

\section{Conclusions}
In this paper the concept of robotic aerial intelligent reflecting surfaces (RA-IRS) is proposed, where an IRS is embedded into a drone which has robotic anchoring capabilities able to grasp in an energy neutral manner at tall urban landforms such as lampposts. By eliminating the energy consumption for hovering and/or flying RA-IRSs can provide services for multiple hours whilst being able to change their anchoring location based on the spatio-temporal characteristics of the traffic demand. To this end, a two-stage optimization framework is proposed to enhance coverage quality whilst minimizing overall energy consumption across a given time period. A wide set of numerical investigations
validated the proposed models and optimization strategies, revealing enhancements, of more than 250\%  on received signal quality compared to nominal fixed deployed IRSs.

\bibliographystyle{IEEEtran}
\bibliography{paper_citation}	

\end{document}